\documentstyle[aps,graphicx,amsmath,amsfonts,amssymb,twocolumn]{revtex}

\def \E {\mbox{E}}
\renewcommand{\vec}[1]{\boldsymbol{#1}}

\newcommand{\var}{\mbox{var\,}}
\newcommand{\norm}[1]{\lVert#1\rVert}
\newcommand{\abs}[1]{\lvert#1\rvert}

\def \a {{\alpha}}
\def \d {\delta \phi}
\def \sg {\sigma}

\def \s {\sin}
\def \c {\cos}

\def \W {{\cal W}}
\def \l {\lambda}

\def \tv {\tilde{\vec{v}}}
\def \tR {\tilde{\vec{R}}}

\def \tlk {t_{\rm lock}}

\def\ltsima{$\; \buildrel < \over \sim \;$}
\def\simlt{\lower.5ex\hbox{\ltsima}}
\def\gtsima{$\; \buildrel > \over \sim \;$}
\def\simgt{\lower.5ex\hbox{\gtsima}}

\begin{document}
\draft

\title{Adaptive filtering techniques for gravitational wave
  interferometric data: Removing long-term sinusoidal disturbances and 
  oscillatory transients.} 

\author{E. Chassande-Mottin$^{*}$ and S. V. Dhurandhar$^{*,\dagger}$}
\address{$^{*}$ Max-Planck-Institut f\"ur Gravitationsphysik,\\
    Albert-Einstein-Institut, Am M\"uhlenberg 1, D-14476 Golm,
    Germany.\\
$^{\dagger}$ IUCAA, Postbag 4, Ganeshkhind, Pune 411 007, India}
\date{\today}

\maketitle

\begin{abstract}
  It is known by the experience gained from the gravitational wave
  detector proto-types that the interferometric output signal will be
  corrupted by a significant amount of non-Gaussian noise, large part
  of it being essentially composed of long-term sinusoids with slowly
  varying envelope (such as violin resonances in the suspensions, or
  main power harmonics) and short-term ringdown noise (which may
  emanate from servo control systems, electronics in a non-linear
  state, etc.). Since non-Gaussian noise components make the
  detection and estimation of the gravitational wave signature more
  difficult, a denoising algorithm based on adaptive filtering
  techniques (LMS methods) is proposed to separate and extract them from
  the stationary and Gaussian background noise. The strength of the
  method is that it does not require any precise model on the observed
  data~: the signals are distinguished on the basis of their
  autocorrelation time. We believe that the robustness and simplicity
  of this method make it useful for data preparation and for the
  understanding of the first interferometric data. We present the
  detailed structure of the algorithm and its application to both simulated
  data and real data from the LIGO 40meter proto-type.
\end{abstract}


\section{Introduction}
\label{intro}
Over the next decade, several large-scale interferometric
gravitational wave detectors will come on-line. These include LIGO,
composed of two Laser Interferometer Gravitational-wave Observatories
situated in the U.S.~\cite{abramovici92:_ligo}, VIRGO, a
French/Italian project located near Pisa~\cite{caron97:_virgo},
GEO600, a German/British interferometer under construction near
Hannover~\cite{lueck97:_geo600}, TAMA in Japan, a medium-scale laser
interferometer \cite{kudora97}, and with funding approval AIGO500, the
proposed 500 meter project sponsored by ACIGA.  There are also
separate proposals for space-based detectors which could be
operational twenty-five years from now (e.g., LISA: the Laser
Interferometer Space Antenna, a cornerstone project of the European
Space Agency \cite{danzmann97:_lisa}). In the meantime, a number of
existing resonant bar detectors will have had their sensitivities
further enhanced.

The key to gravitational wave detection is the very precise
measurement of small changes in distance. For laser interferometers,
this is the distance between pairs of mirrors hanging at either end of
two long, mutually perpendicular vacuum chambers. Gravitational waves
passing through the instrument will shorten one arm while lengthening
the other. By using an interferometer design, the relative change in
length of the two arms can be measured, thus signaling the passage of
a gravitational wave at the detector site.  Long arm lengths, high
laser power, and extremely well-controlled laser stability are
essential to reach the requisite sensitivity, since the gravitational
waves will be faint and will modify only weakly the structure of
space-time in the detector's arms (see e.g.,
\cite{saulson94:_fundam_gw_detec}).

Gravitational wave detectors produce an enormous volume of output
(e.g., of the order of 16 MB/sec for the LIGO instruments) consisting
mainly of noise from a host of sources both environmental and
intrinsic to the apparatus. Buried in this noise will be the
gravitational wave signature. Sophisticated data analysis techniques
will have to be developed to optimally extract astrophysical data.
Many of the techniques developed so far~
\cite{allen99:_obser_galax,mohanty98:_hierar,sathyaprakash91:_choic} are based on
matched filtering and assume stationary Gaussian noise.

However, the real data stream from the detectors is not expected to
satisfy the stationary and Gaussian assumptions. In fact, the data
from the Caltech 40 meter proto-type interferometer has the expected
broadband noise spectrum, but superposed on this are several other
noise features~\cite{allen99:_obser_galax}; such as long-term
sinusoidal disturbances emanating from suspensions and electric main
harmonics and also transients occurring occasionally, typically due to
servo-controls instabilities or mechanical relaxation in suspension
system etc.  While no precise \textit{a priori} model can be given for
this noise until the detector is completed and fully tested, matched
filtering techniques cannot be used to locate/remove these noisy
signals.

This disparity between standard Gaussian assumptions and real data
characteristics poses a major problem to the direct application of
matched filtering techniques. This is true when searching for burst
sources such as blackhole binary quasinormal ringings
\cite{creighton97:_listen}. This is also the case for the inspiral
searches in Caltech 40meter data, where one has to introduce a veto
\cite{allen99:_obser_galax} on the decision taken with the matched
filter to ensure that the detected signal is actually the one we are
looking for.

It is possible that in the future, improved experimental
techniques and greater experience, will reduce or even completely 
eliminate some of these nonstationary and non-Gaussian features.
Nevertheless, it will take probably some time to reach such acceptable
and high quality of data. Therefore, it is necessary and desirable to 
somehow combat this noise. Since such noise features defy modeling, a 
novel approach to the problem is called for.

We propose a denoising method based on \textit{LMS adaptive linear
  prediction} techniques which does not require any precise \textit{a
  priori} information about the noise characteristics. Although our
method does not pretend to optimality, we believe that its simplicity
makes it useful for data preparation and for the understanding of the
first data.

In the following, we present the principles of LMS adaptive denoising
(Sect. \ref{method}), a characterization of its behavior on a simple
model of the noise from the interferometer (Sect. \ref{aleforgw}), the
precise structure of the denoising algorithm (Sect.
\ref{aleinpractise}) and results (Sect. \ref{results}) obtained
with simulated data and also with real data taken from the Caltech 40
meter proto-type interferometer~\cite{abramovici96:_improv_ligo}.

This work here is preliminary; its goal is to explore how effectively 
adaptive filtering techniques perform on the problem we 
address.  It is a first step towards a more complete statistical
evaluation of the algorithm.

\section{Methods}
\label{method}
\subsection{From hypothesis to method}
We assume that the noise consists of broadband Gaussian noise plus
large amplitude oscillating interference signals. The model does not
include any \textit{a priori} knowledge of the signal such as its
exact frequency or shape of the envelope. The only assumption we make
is that its autocorrelation over a small time-lag -- the time-lag
chosen greater than the decorrelation time scale of the broadband
noise -- is appreciable, while for the broadband noise it is
essentially zero.  This difference can be used to advantage to
discriminate between the narrow band interferences and the broadband
noise.

The idea is to predict the current signal sample given the previous
samples of the data. This is possible, only if the target
sample shares enough information with (i.e., is sufficiently
correlated to) the previous samples. In other words, the only predictable
part of the signal is the one whose correlation length is sufficiently
large (i.e., long-term sinusoids or ringdowns). Conversely, 
broadband noise cannot be predicted, as it is not possible to guess
the next value in this way.  It is this crucial underlying idea we use
to discriminate between the two noise signals.

\subsection{Mean square linear prediction}
Let us recall some standard principles to design an optimal linear
predictor. The question to address is to optimally predict the data
sample $x_k$ with a collection of past samples $\vec{x}_k: =
(x_{k-d-m}, m = 0, 1,\ldots, N-1)^{t}$ given the delay $d \geq 1$ (the
quantity $d$ is also referred to as prediction depth). The prediction
is obtained by linearly combining these data samples weighted by the
$N$ corresponding coefficients\footnote{We put brackets around indices
  of vectors and matrices in order to distinguish them from the time
  index.} $w^{(m)}$, forming the tap-weight vector $\vec{w}:=(w^{(m)},
m = 0, 1,\ldots, N-1)^{t}$, where the superscript `t' denotes the
transpose of the vector.  Therefore, the prediction $y_k$ of $x_k$
reads,
\begin{equation} 
y_k:=\vec{w}^t \vec{x}_k.  
\end{equation}

The predictor is optimal in the mean square sense when the variance of
the prediction error $e_k=x_k-y_k$ is minimum. Therefore, the problem
is to find the set of weight coefficients which minimizes
\begin{equation} 
\label{varerror1}
J_k(\vec{w}):=\E[e^2_k]=\E[(x_k-\vec{w}^t \vec{x}_k)^2],
\end{equation}
where $\E$ denotes the expectation value operator.

This leads to the minimization of the following quadratic form
\begin{equation}
\label{varerror2}
J_k(\vec{w})=\sigma^2_k-2 \vec{w}^t \vec{p}_k+\vec{w}^t \vec{R}_k \vec{w},
\end{equation}
where $\sigma^2_k:=\E[x^2_k]$, $\vec{p}_k:=\E[x_k \vec{x}_k]$ and
$\vec{R}_k:=\E[\vec{x}_k^t \vec{x}_k]$.  There exists only one solution
$\vec{w}^*_k$, obtained when the gradient of $J_k$ vanishes. This
situation is realized when
\begin{equation}
\label{wiener}
\vec{R}_k \vec{w}^*_k= \vec{p}_k.
\end{equation}

When the signal is stationary, $\vec{R}_k=\vec{R}$ and
$\vec{p}_k=\vec{p}$ are constant (independent of $k$). In this case,
$\vec{R}$ defines the autocorrelation matrix of the signal $x_k$ and
the solution of (\ref{wiener}) is referred to as the \textit{Wiener
  filter}.

\subsection{Linear prediction and LMS method}
Eq. (\ref{wiener}) requires the computationally expensive inversion of
the matrix $\vec{R}_k$. An alternative and more efficient solution
for finding the minimum of the (\ref{varerror2}) consists in starting
from an arbitrary initial value $\vec{w}_{k,0}$, and iterate the
tap-weight vector along the steepest descent direction,
\begin{equation}
\label{recurs}
\vec{w}_{k,n+1}=\vec{w}_{k,n} - \mu \nabla_{\vec{w}} J_k(\vec{w}_{k,n}),
\end{equation}
given by the gradient
\begin{equation}
\label{gradj1}
\nabla_{\vec{w}} J_k(\vec{w})=2(\vec{R}_k \vec{w}-\vec{p}_k). 
\end{equation}

For a sufficiently small gain $\mu$, the weight vectors will
eventually converge to the optimal predictor filter $\vec{w}_k^*$.
This procedure requires the second order statistics (namely
$\vec{R}_k$ and $\vec{p}_k$) of the signal. In our case, this 
information is not available and one has therefore to estimate these
quantities. Instead of estimating directly $\vec{R}_k$ and $\vec{p}_k$
and combining them with (\ref{gradj1}), a more efficient solution is
to estimate the gradient.  From the derivation of (\ref{varerror1}),
one can rewrite the gradient as
\begin{equation}
\label{gradj2}
\nabla_{\vec{w}} J_k(\vec{w})=-2\E[e_k \vec{x}_k].
\end{equation}

A simple and natural way to obtain an estimator of this quantity is
to omit the expectation operator~:
\begin{equation} 
\label{gradlms}
\widehat{\nabla_{\vec{w}} J_k}=-2 e_k \vec{x}_k.
\end{equation} 

Because the noise perturbs this estimate, the algorithm may
iterate in a direction which does not lie along the direction of 
steepest descent, thus preventing the filter from converging to the Wiener
filter. For this purpose, we stabilize the estimation above by setting
the algorithm time index $n$ equal to signal time index $k$ in the Eq.
(\ref{recurs}). The final evolution equation for the tap-weight vector
finally reads:~
\begin{equation}
\label{lmscoef}
\vec{w}_{k+1}=\vec{w}_{k}+ 2\mu e_k \vec{x}_k.
\end{equation}

At a fixed time $k$, the weight vector evolves along the crude estimate
of the steepest descent direction. But on a longer duration, the
direction followed by the tap-weight vector is governed by the
sum of the successive gradient estimates obtained with
different noise samples. In other words, we have replaced an ensemble
average in (\ref{gradj2}) by a time average. It also implies that we have
implicitly called for further assumptions on the signal $x_k$: first
its local stationarity (more precisely, the second order statistics
are supposed to be constant during the convergence time of the
algorithm) and second, its ergodicity.

\begin{figure}
\begin{center}
\setlength{\unitlength}{2000sp}
\begin{picture}(10244,4161)(579,-3908)
\thicklines
\put(8401,-361){\circle{1200}}
\put(601,-361){\line( 1, 0){7200}}
\put(1501,-361){\line( 0,-1){2400}}
\put(1501,-2761){\line( 1, 0){900}}
\put(9001,-361){\line( 1, 0){1800}}
\put(7501,-2761){\line( 1, 0){900}}
\put(8401,-2761){\line( 0, 1){1800}}
\put(5701,-3361){\framebox(1800,1200){}}
\put(2401,-3361){\framebox(1800,1200){delay}}
\put(4201,-2761){\line( 1, 0){1500}}
\put(8101,-361){\line( 1, 0){600}}
\put(10001,-361){\line( 0,-1){3525}}
\put(10001,-3886){\line(-1, 0){2800}}
\put(7201,-3886){\vector(-1, 2){1200}}
\put(1,-61){\makebox(0,0)[lb]{input signal, $x_k$}}
\put(9500,-61){\makebox(0,0)[lb]{error, $e_k=x_k-y_k$}}
\put(6400,-3100){\makebox(0,0)[lb]{$\vec{w}_k$}}
\put(8626,-1411){\makebox(0,0)[lb]{\footnotesize prediction,}}
\put(8626,-1811){\makebox(0,0)[lb]{$y_k$}}
\put(5750,-2700){\makebox(0,0)[lb]{\footnotesize adaptive filter}}
\end{picture}
\end{center}
\caption{\label{schema} The figure illustrates the principle of the
  underlying method on which the algorithm we propose is based. The algorithm is 
  designed to discriminate the nonstationary and non-Gaussian noise 
  features from the broadband background noise in interferometric 
  gravitational wave data.  This method is referred to as 
  \textit{LMS adaptive line  enhancement} and its objective is to compare 
  the signal and its linear prediction, the predictor coefficients
  being adjusted by a feedback loop controlled by the prediction error.}
\end{figure}
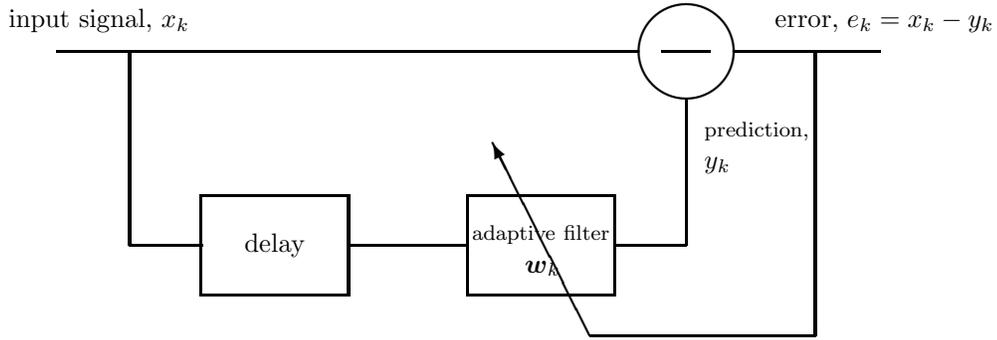

Summarizing, the method we propose consists in linearly filtering the
data to extract the part of the signal with a long correlation
time. As illustrated with the block diagram in Fig. \ref{schema},
the finite impulse response filter (given by $\vec{w}_k$) is modified
at each iteration according to the relation (\ref{lmscoef}) with the
final goal to minimize the mean square error. Once the filter has
converged (i.e., $\vec{w}_k$ is stable in time), we reject the
predicted part of the signal (corresponding to the long-term
sinusoidal or the ringdown signals) and we send the rest of the signal
for further analysis for detection.

\subsection{Properties of the LMS method}
The method we described above is referred to as \textit{adaptive line
  enhancer (ALE)}.  It is a special case of the \textit{LMS
  algorithm}. Both, ALE and LMS algorithms have been first introduced
by Widrow and Hoff \cite{widrow84:_adapt} in the 1960's.

The acronym LMS (Least Mean Square) designates a general scheme to
design signal processing methods where a minimization (in a
statistical sense) of a definite positive quadratic cost function
(usually related to some mean quadratic error) is needed. Its central
idea is the use of the estimate of the gradient of this function given
in Eq. (\ref{gradlms}). The LMS technique has been extensively used
for the last 30 years in communications problems such as echo
cancellation, channel equalization, antenna processing, etc. The main
advantages to be gained by applying the LMS technique are (\textit{i})
adaptivity, (\textit{ii}) robustness, (\textit{iii}) simplicity.

In this context, the term ``adaptivity'' has two different meanings.
First, it means that the LMS technique will automatically modify its
parameters to reach for the best setup for a problem which has not
been initially precisely defined. Second, it is also able to follow changes
in the characteristics of the data being processed in the event that 
they occur. The latter property also shows that the method is robust. 
In fact, this
method has been proved to be robust according to specific statistical
criterion such as the minimax criterion \cite{haykin96:_adapt}.

The ALE is an adaptive prediction algorithm using the LMS technique.
We have seen that the signal is predicted from a reference signal
which is the signal itself. In some other applications, although the same
principles are applied, the reference signal can be another signal,
e.g. echo cancellation or denoising. In such cases, the quantity of
interest might not be the prediction output but the linear filter used
to compute it, e.g., deconvolution.

\section{Adapting ALE filter to canceling noise in GW data}
\label{aleforgw}

In this section we essentially describe a model for understanding the
behaviour of the ALE algorithm. The model we assume consists of a high
amplitude narrowband signal superposed on broadband noise. For
simplicity, we assume the broadband noise to be white and Gaussian and
the narrowband signals are sinusoids of constant envelope. The results
we obtain hold for more realistic signals when the evolution of
their amplitude and/or instantaneous frequency occurs adiabatically,
i.e., the change is small over the period of the sinusoid.
%We will consider subsequently more realistic signals with slowly
%varying envelope.

The assumption of white noise is not too restrictive because this is
equivalent to choosing the noise correlation time to be zero and
therefore we are free to choose the prediction depth (i.e., the time
delay between the current predicted data sample and the reference
signal to the LMS filter) to be arbitrarily small. In a real
situation, we must fix the delay to be greater than the correlation
time of the broadband noise. We first analyze the case of the sinusoid
because it is easier to investigate and provides invaluable insights
into the workings of the LMS algorithm.

It may be remarked that the denoising of sinusoids in white noise has
been treated in the literature with great detail (see
\cite{haykin96:_adapt,zeidler90:_perfor_lms} for a review). We give
here only pertinent results (with a short proof) for introducing the
structure of the algorithm, which we present later in the text.

%\subsection{The example of the sinusoid}
\label{sn}
\subsection{Optimal filter}
We consider the data to be of the following form,
\begin{equation}          
\label{signaltest}
x_k :=  \c (2 \pi f_0 t_k+\Phi) + n_k,
\end{equation}
where $t_k:= k \delta$, $\delta:=1/f_s$ being the sampling interval and
$\Phi$ is a random phase (at the origin) with uniform probability
density function between $-\pi$ and $\pi$. The sinusoid has frequency
$f_0$ and the units are so chosen that it is of unit amplitude. The
additive white noise $n_k$ with variance $\sg^2$ satisfies the
relation,

\begin{equation}
\E[n_k n_m] := \sg^2 \delta_{km},
\end{equation}
where $\delta_{km}$ is the Kronecker delta.

The reference signal to the adaptive filter is just the delayed data
by the amount $d \delta$, where $d$ is the number of time samples.  We
choose $N$ weights $w_k \equiv W$ ($\vec{w}$ can be thought of as a
column vector) for the length of our filter, then the ``reference
vector'' $\vec{x}_k$ at the $k$th time instant $t_k$ has the
components $x_{k-d-n}$, $n = 0, 1,\ldots, N-1$. The components of the
autocorrelation matrix $\vec{R}$ and the vector $\vec{p}$ in
Eq.(\ref{varerror2}) are given by
\begin{align}
\label{autcor}
\vec{R}^{(mn)} &= 1/2 \cos(m-n) \d + \sg^2 \delta_{mn}\\
\vec{p}^{(m)} &= 1/2 \cos(m+d) \d,
\end{align}
where $(m,n)=0, 1, \ldots, N-1$ and $\d = 2 \pi f_0 \delta$. Note that
we have dropped the index $k$ because the autocorrelation $\vec{R}$
does not depend upon $k$, since we are dealing with a stationary
signal.

From the above expressions of $\vec{R}$ and $\vec{p}$ and solving 
Eq.(\ref{wiener}), we obtain the optimum Wiener filter
\begin{equation}
\label{wstar}
\vec{w}^{* (m)}=\frac{2}{N + 4 \sg^2}\,\cos(m+d)\d, \quad m=0, 1,\ldots, N-1.
\end{equation}
where we have chosen the length of the filter to be half-integral 
number of cycles for reasons of simplicity, i.e. $N \d = l \pi$, 
where $l$ is an integer.

In other words, the optimum linear predictor is nothing but a copy of
the expected signal itself. The filter in Eq.  (\ref{wstar}) is also
referred to as matched filter. In our situation, in practise, 
$N \gg \sg^2$ and the term $4 \sg^2$ can be omitted from the amplitude 
of $\vec{w}^*$. 

For the reasons detailed before, we propose to use the ALE algorithm
in order to find a good approximation of $\vec{w}^*$.  Starting from
an arbitrary initial tap-weight vector, we iterate the weights
$\vec{w}_k$ according to Eq. (\ref{lmscoef}) to converge to
$\vec{w}^*$. Once the filter is ``close'' enough to the optimal
solution (the word ``close'' will be defined later in the text), we
then say that the filter has locked on to the signal.

\subsection{Approach to locking}
\label{approachtolock1}
\paragraph{Continuous time approximation of the locking trajectory ---}
We may analyze the approach to locking by deriving a difference
equation for the averaged evolution of the weights and then investigating
this equation. It is impossible to obtain the average evolution of the
weights by using the standard definition of the expectation operator
$\E$ because of the nonlinearity and the recursive scheme involved in
evolving the weights. We therefore adopt the time-average over
successive data points as the operational definition of $\E$.

Shifting the origin to $\vec{w}^*$ by defining $\vec{v}_k := \vec{w}_k
- \vec{w}^*$, we may write the LMS evolution equation (\ref{lmscoef})
in the following form \cite{macchi95:_adapt}:
\begin{equation} 
\label{evol}
\vec{v}_{k+1}-\vec{v}_{k} = - 2\mu (\vec{x}_k \vec{x}_k^t)\vec{v}_k +
2\mu e_k^* \vec{x}_k,  
\end{equation} 
where $e_{k}^{*}:=x_k-\vec{w}^{*t}\vec{x}_k$ is the prediction error
produced when using the optimal filter.

During the locking phase, the filter is far apart from the optimal
location (i.e., $\vec{v}_k$ has a large modulus). The homogeneous term
dominates the forcing term in the difference equation (\ref{evol})
which then can be approximated by~:
\begin{equation}
\vec{v}_{k+1}-\vec{v}_{k} = - 2\mu (\vec{x}_k \vec{x}_k^t)\vec{v}_k.
\end{equation} 

In the situation where the step gain parameter $\mu$ is chosen to be
very small so that the weight coefficients are almost constant over a
given time interval, the recursivity eventually acts as an averaging
operation on both sides of the equation above. This leads to the
difference equation which we use to describe the tap-weight trajectory
in the space of weight coefficients, we denote $\W$~:
\begin{equation} 
\label{evol_approach}
\vec{v}_{k+1}-\vec{v}_{k} = - 2\mu \vec{R} \vec{v}_k.  
\end{equation}

Let $\vec{Q}$ be the transformation which diagonalizes $\vec{R}$. The
above difference equation is best analyzed by changing the frame in
$\W$ to the principal axis
\begin{equation} 
\label{eigevol}
\tv_{k+1} - \tv_k= - 2 \mu \tR \tv_k,
\end{equation}
where $\tR:=\vec{Q}\vec{R}\vec{Q}^{-1}={\rm diag}(\lambda^{(0)},
\ldots, \lambda^{(N-1)})$ and $\tv_k := \vec{Q} \vec{v}_k$.  Eq.
(\ref{eigevol}) gives decoupled difference equations for the components
$\tv^{(m)}_k, m = 0,\ldots,N-1$ of $\tv$ which can be solved given the
initial weight vector $\tv^{(m)}_0$~:
\begin{equation}
\label{wt}
\tv^{(m)}_k  = \tv^{(m)}_0 (1 - 2 \mu \l^{(m)})^k.
\end{equation}

\paragraph{Eigenvalues and eigenvectors of $\vec{R}$ ---}
We need to compute the eigenvalues $\l^{(m)}$ of $\vec{R}$. This can be
conveniently implemented by splitting $\vec{R}$ into the noise part, which is
just $\sg^2$ times the identity plus the signal part which we denote
by $\vec{S}/2$, and thus,
\begin{equation}
\label{aut}
\vec{R} = \sg^2 \vec{I} + \vec{S}/2,
\end{equation}
where $\vec{S}^{(mn)}:=\c (m-n)\d$. It is easily verified that the
eigenvectors of $\vec{R}$ and $\vec{S}$ are identical and the
eigenvalues of $\vec{R}$ are obtained from those of $\vec{S}$ by first
halving them and then adding $\sg^2$ to the result. It remains,
therefore, to compute the eigenvalues and eigenvectors of $\vec{S}$.
We do this by observing that we can write $\vec{S}$ as follows
\footnote{By definition, the vectors $\bar{\vec{x}}$ and
  $\vec{x}^{\dagger}:=\bar{\vec{x}}^t$ denote respectively the complex
  conjugate and the hermitian transpose of $\vec{x}$.}:
\begin{equation}
\label{sig}
\vec{S}=(\vec{\nu}\vec{\nu}^{\dagger}+\bar{\vec{\nu}}\bar{\vec{\nu}}^{\dagger})/2,
\end{equation}
where $\vec{\nu}:=(1, \exp (i \d ), \exp (2i\d),\ldots,
\exp((N-1)i\d))^t$. 

Since the matrix $\vec{S}$ is real and is essentially made out of two
external products of $\vec{\nu}$ and $\bar{\vec{\nu}}$, its rank 
equals $2$ ($\vec{S}$ has $N-2$ degenerate eigendirections in $\W$ with
eigenvalue zero) with two non-zero real eigenvalues. Let $\vec{v}$ be
an eigenvector associated to one of the non-trivial eigenvalues.
According to the structure of $\vec{S}$, the vector $\vec{v}$ can be
written without loss of generality as the following linear
combination, 
\begin{equation}
\label{vdef}
\vec{v}= \vec{\nu}\,\exp(-i\a) + \bar{\vec{\nu}}\,\exp (i\a),
\end{equation}
where the coefficients have been chosen to have unit modulus arbitrarily.

Using the two scalar products $\vec{\nu}^{\dagger} \vec{\nu} = N$
and
\begin{align}
\vec{\nu}^t\vec{\nu}&=1+\exp(2i\d)+\ldots+\exp(2(N-1)i\d)\\
&:=\beta\exp i\gamma,
\end{align}
where the geometric series can be summed up and modulus and phase ascertained~:
\begin{align}
\label{betagamma}
\beta&= \frac{\s(N\d)}{\s\d}\\
\gamma&=(N-1)\d,
\end{align}
we obtain the effect of the matrix $\vec{S}$ on the vector $\vec{v}$, given
by
\begin{equation}
\label{sv}
\vec{S}\vec{v} = \vec{\nu}\,\exp (- i \a) (N + \beta \exp (- i
\gamma + 2 i \a))/2 + c.c.
\end{equation}
where \textit{c.c.} denotes complex conjugate.

This expression has to be compared to the second term of the
eigenvalue equation $\vec{S}\vec{v}=\l \vec{v}$, leading to two
solutions for $\a$, namely, $\a=\gamma/2$ and $\a=(\gamma - \pi)/2$.
These yield the eigenvectors $\vec{v}_{\pm}$ and the corresponding
eigenvalues $\l_{\pm}$~:
\begin{equation}
\label{eigvec}
\begin{array}{rcl}
\vec{v}_{+} &=& \vec{\nu} \exp (- i\gamma/2)+ \bar{\vec{\nu}} \exp(i\gamma/2),\\[2mm]
\vec{v}_{-} &=& i \vec{\nu} \exp(-i\gamma/2) - i \bar{\vec{\nu}} \exp(i\gamma/2),
\end{array}
\end{equation}
\begin{equation}
\label{eigval}
\l_{\pm} = (N \pm \beta)/2.
\end{equation}

If we choose $N$ large enough and $N\d = m \pi$, where $m$ is an
integer then the analysis becomes simpler. This amounts to choosing
the length of the filter to have half-integral number of cycles: we
have $\beta = 0$ and $\l_{\pm} = N/2$. (Geometrically, this means that
the eigenvalue problem is degenerate with respect to the two signal
eigenvectors: there is a two dimensional eigenspace belonging to the
eigenvalue $N/2$. The weights thus evolve non-preferentially with
respect to the signal eigendirections.) Since typical cases imply
generally $N \gg \beta$, we will assume this simplification in the
rest of the paper.

In this situation, the spectrum of $\vec{R}$ 
\begin{multline}
\label{rspectrum}
\mbox{sp}(\vec{R})=\{\l^{(0)}=\l^{(1)}=N/4+\sg^2 \quad
\text{ and }\\
\l^{(m)}=\sg^2, m=2,\ldots,N-1\},
\end{multline}
consist of two sets of eigenvalues~: the first two correspond to
directions in the signal space associated to ``signal+noise'' (or
``signal'', for short) whereas the remaining $N-2$ characterize
``noise'' directions.

According to Eq. (\ref{wt}), the weight vector will converge more
rapidly in directions associated with the largest eigenvalues, which
are the signal eigenvalues. The other noise eigenvalues are
unimportant in this consideration. The eigenvectors pertaining to the 
signal provide preferred directions in
$\W$: it is along these directions that the slope of the performance
surface is steep and hence promotes faster convergence.

\subsection{Steady state evaluation}
\label{steadystate}
If the step gain factor is sufficiently small, the tap-weight coefficients
eventually converge and stabilize in a neighbourhood of the optimal
value. At this stage, the assumptions made in obtaining the
approximated evolution equation (\ref{evol_approach}) do not hold
anymore. In contrast with the case of ``the approach to locking'', 
the right hand side of the
difference equation (\ref{evol}) is now dominated by the forcing
term~:
\begin{equation} 
\label{evol_steady}
\vec{v}_{k+1}-\vec{v}_{k} = 2\mu e_k^* \vec{x}_k.  
\end{equation} 

Roughly speaking, the trajectory of the vector $\vec{w}_k$ during the
steady state can be viewed as a random walk centered around $\vec{w}^{*}$
lying within a region of $\W$ space whose extent is 
determined by two factors, namely, $\mu$ and the intrinsic geometry of 
$\W$ in the vicinity of $\vec{w}^{*}$.

The misalignment between the actual ALE filter $\vec{w}_k$ and the
optimal one $\vec{w}^*$ creates an additional error in the output.
In fact, a direct calculation from Eq. (\ref{varerror2}) shows that
the total mean square error may decomposed as \cite{widrow84:_adapt}~:
\begin{equation}
\label{mse}
J_k(\vec{w}_k)= \xi_{min}+\xi_k,
\end{equation}
where (\textit{i}) $\xi_{min}:=J_k(\vec{w}^*)$ is the \textit{minimum mean square
  error} arising from the fraction of the input noise which still
remains in the output, assuming that the ALE filter has reached
exact optimality and (\textit{ii}) $\xi_k:=\vec{v}_k^t \vec{R} \vec{v}_k$ is the
\textit{excess mean square error} (EMSE) due to the misalignment
between the ALE filter and the Wiener filter.

One can verify that the EMSE vanishes when reaching optimality i.e.,
when $\vec{v}_k=\vec{0}$. In other words, this term quantifies the non
optimality of the current filter in use. We can imagine $\xi_k$ as the
square of a natural distance in $\W$ and $\vec{R}$ as a intrinsic
metric over $\W$.

A good approximation of $\xi_{min}$ can be found for large number
of weight coefficients for the specific case of sinusoidal signals
with high SNRs.
Using Eqs. (\ref{varerror2}) and (\ref{wiener}), we may write
$\xi_{min}=\E[x_k^2]-\vec{p}^t \vec{w}^{*}$.  When $N \rightarrow
\infty$, a direct calculation shows that the second term $\vec{p}^t
\vec{w}^{*}$ tends to the energy of the sinusoid, which means that 
the remaining energy is that of the noise:~
$\xi_{min} \approx \sigma^2$.

We complete the characterization of the mean square error (\ref{mse})
with the evaluation of the average value of the EMSE, which we denote
by $\xi^{(st)}$. Firstly, noticing that the EMSE is invariant under the
principal axis transformation
\begin{equation}
\label{err}
\xi_k=\tv_k^t \tR \tv_k = \sum_{m=0}^{N-1} \l^{(m)} \left(\tv^{(m)}_k\right)^2,
\end{equation}
and secondly, using the approximation $\E[\tv_k \tv_k^t] \approx \mu
\xi_{min} \vec{I}$ proposed in \cite{widrow84:_adapt} to obtain the
typical value for $(\tv^{(m)}_k)^2$ yields
\begin{equation}
\xi^{(st)} \approx \mu \sum_{m=0}^{N-1}{\l^{(m)}} \xi_{min}.
\end{equation}

Since the signal is of much larger amplitude than the broadband 
noise, the trace of $\tR$ is essentially due to the signal eigenvalues
(see Eq. \ref{rspectrum}). Combining with the expression of
$\xi_{min}$ above, this leads to~:
\begin{equation}
\label{emse}
\xi^{(st)} \approx \mu N \sigma^2/2.
\end{equation}

A better estimate of $\xi^{(st)}$ can be obtained starting with more realistic
hypotheses and using more sophisticated approximations 
\cite{zeidler90:_perfor_lms}~:
\begin{equation}
\xi^{(st)} \approx \sum_{m=0}^{N-1}\frac{\mu \l^{(m)}}{1 - \mu
\l^{(m)}} \xi_{min}.
\end{equation}

In the limit of small step size, this approximation tends to the
simpler one in Eq. (\ref{emse}).

\subsection{Convergence time}
In the expression of the EMSE in Eq. (\ref{err}), we separate the sum
into two parts: the first, $\xi^{(n)}_k$ associated with the noise
(i.e., consisting of terms involving noise eigenvalues and vectors),
the second, $\xi^{(s)}_k$ with the signal. Because the signal
eigenvalues are much larger than those of the noise, the sum in Eq.
(\ref{err}) is essentially dominated in the beginning (for small $k$)
by $\xi^{(s)}_k$. These two errors decrease during the locking phase
until reaching a steady state value. The locking time (i.e., time at
which the steady state is reached) is defined to be that, when
$\xi^{(s)}_k$ is of the order of the total EMSE expected in the 
steady state.

From Eqs. (\ref{wt}), (\ref{eigval}) and (\ref{err}) we obtain,
\begin{align} 
\label{emsesig}
\xi^{(s)}_k&:=\l^{(0)} \left(\tv^{(0)}_k\right)^2+\l^{(1)} \left(\tv^{(1)}_k\right)^2\\
&\approx \frac{N}{4}
\left(\left(\tv_0^{(0)}\right)^2 + \left(\tv_0^{(1)}\right)^2\right)
\left(1 - \frac{\mu N}{2}\right )^{2 k}, 
\end{align}
where we have assumed $N \gg \beta$. 

We set the starting point $\vec{w}_0$ in $\W$ to be $\vec{0}$. It
corresponds to the initial value $\tv_0$ in the eigenspace which is
given by 
$\tv_0=-\vec{Q}\vec{w}^*$. The first two coordinates of $\tv_0$ can be
directly obtained since the first two row vectors of $\vec{Q}$ are
just the normalized signal eigenvectors of the $\vec{R}$ matrix,
leading to, for half integral wavelength filters and $N \gg \sg^2$,
\begin{align} 
\tv^{(0)}_0&= \sqrt{2/N} \cos (d\d +\gamma/2)\\ 
\tv^{(1)}_0&= \sqrt{2/N} \sin (d\d + \gamma/2).
\end{align}

These considerations yield
\begin{equation}
\xi^{(s)}_k \approx \frac{1}{2}\left(1-\frac{\mu N}{2}\right)^{2k}.
\end{equation}

This error must now be compared with the averaged EMSE in Eq. (\ref{emse})
in order to find the time $\tlk$ at which $\xi^{(s)}$ and $\xi^{(st)}$
are equal~:
\begin{equation} 
\label{tlock}
\tlk \approx  \delta \frac{\ln(\mu N \sg^2)}{2 \ln(1 - \mu N/2 )}. 
\end{equation} 

It is important to mention that, when the product $\mu N/2$ tends to
$1$, the convergence time diverges to infinity meaning that the
weights do not converge toward $\vec{w}^{*}$ anymore. In order to
ensure the stability of the algorithm, the parameters will have to
satisfy the stability condition $0 < \mu N/2 < 1$. However, we have
observed in our simulations that when $1/2 \leq \mu N/2 <1$, the
convergence is slowed down, because of the presence of oscillatory
terms in the gradient which do not average to zero anymore. In
practise, it is advisable to choose the parameters so that $\mu N/2 <
1/2$. 

For a sinusoid of amplitude $A$ instead of unity as we have considered
before, the condition for stability can be simply obtained by
replacing the parameter $\mu N/2$ by $\rho:=\mu N A^2/2$ leading to $0
< \rho < 1$.

\begin{figure}
  \centerline{\includegraphics[height=100mm]{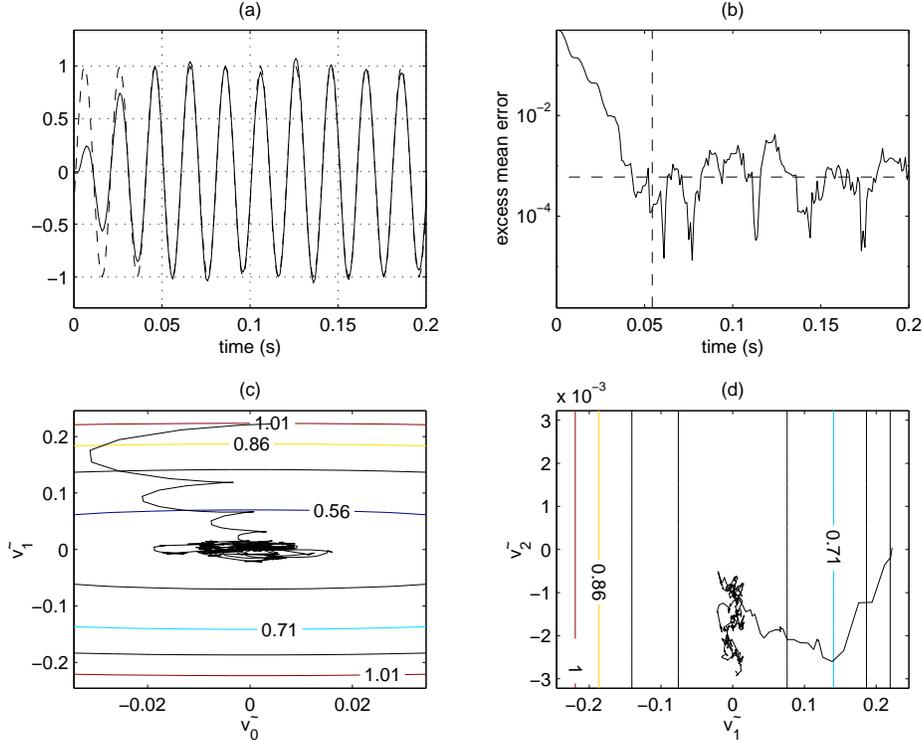}}
  \caption{\label{test1}\textbf{Applying the ALE to a sinusoidal
   signal: approach to locking and steady state}. The figure illustrates how
    the ALE performs on a test signal composed of a sinusoid ($f_0=50$
    Hz sampled at $f_s=1$ kHz) corrupted by white noise
    (SNR$=A^2/(2\sigma^2)=50 (17 \mathsf{dB})$). We initialize the
    $N=40$ tap-weight coefficients to $0$, set the prediction
    depth $d=5$ (ms) and the step gain parameter $\mu=0.003$.
    (\textbf{a}): the filter output signal $y_k$ (solid line)
    converges rapidly towards the actual noise free sinusoidal signal
    (dashed line). (\textbf{b}): this is confirmed by observing that the
    ``signal'' EMSE $\xi^{s}_k$ defined in Eq. (\protect\ref{emsesig})
    which decreases with time until it reaches its steady state value.
    For comparison, the horizontal dashed line indicates the
    theoretical mean value of the total EMSE $\xi^{st}$ in the steady
    state (see Eq. (\protect\ref{emse})). We can verify that the
    theoretical value obtained in Eq.  (\protect\ref{tlock}) for the
    convergence time $t_{lock}$ corresponds effectively to the time
    instant at which $\xi^{s}_k$ and $\xi^{st}$ are of the same order.
    Finally, the two contour plots of the bottom line display the
    trajectory followed by the adaptive filter coefficients in the
    eigen weight space~: in \textbf{(c)}, the axis are the first two
    eigenvectors of $\vec{R}$ namely $\tilde{\vec{v}}_k^{(0)}$ and
    $\tilde{\vec{v}}_k^{(1)}$ (i.e., the ``signal'' eigenvectors)
    whereas in \textbf{(d)} the diagram plane is given by
    $\tilde{\vec{v}}_k^{(1)}$ and $\tilde{\vec{v}}_k^{(2)}$ (i.e., a
    ``signal'' direction vs. a ``noise'' direction). As proved in
    Sect.  \protect\ref{approachtolock1}, the weight coefficients
    converge more rapidly along the two directions
    $\tilde{\vec{v}}_k^{(0)}$ and $\tilde{\vec{v}}_k^{(1)}$ given by
    eigenvectors associated with the largest eigenvalues.}  
\end{figure}

\begin{figure}
  \centerline{\includegraphics[width=110mm,height=50mm]{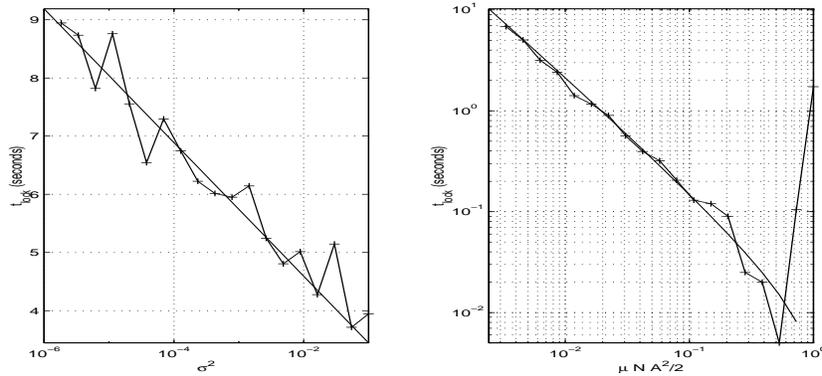}}
  \caption{\label{test1b} \textbf{Applying the ALE to a sinusoidal
      signal: convergence time}. The figure shows the comparison
    between the convergence time $t_{lock}$ obtained by 
    simulations (solid line with '+' associated with one noise
    realization) and its theoretical value (solid line) given in Eq.
    (\protect\ref{tlock}). The test signal is a sinusoid in white
    Gaussian noise (see Eq. (\protect\ref{signaltest}) with $A=1$,
    $f_0=50$ Hz, sampled at $f_s=200$ Hz). The convergence time is
    shown as a function of $\sigma^2$ in (\textbf{a}) (we have fixed
    the remaining ALE parameters to $N=200$ and $\mu N A^2/2=0.01$)
    and in (\textbf{b}) as a function of $\mu N A^2/2$ ($N=200$ and
    $\sigma^2=0.1$). Simulations globally confirm the results obtained
    in Eq. (\ref{tlock}) except when $1/2 \leq \mu N A^2/2< 1$. 
    The reason for this discrepancy is that the difference
    between the actual gradient (\protect\ref{gradj2}) and its
    estimates (\protect\ref{gradlms}) can be shown to be an
    oscillating term which does not average to zero any longer when the
    step gain parameter approaches the critical value for stability.
    Therefore, in practise, we choose parameters so that $\mu N A^2/2 <
    1/2$.}
\end{figure}

\vspace{5mm} 
We illustrate in Figs. \ref{test1} and \ref{test1b} with an example
the results of this Section pertaining to the approach to locking
and steady state analysis.

\section{The ALE in practice}
\label{aleinpractise}
In the previous Sections we have characterized the behaviour of the
ALE in cases of interest. We will now elaborate on how this algorithm can
be adapted to the interferometric data. 
 
In the scheme we present here, we first decompose the signal in $p$
frequency subbands to which we apply the ALE twice with different sets
of parameters. In the first stage, the parameters are tuned to best
remove long-term sinusoidal components of the noise; whereas in the
second stage, the target consists of shorter oscillatory transients.

\subsection{Subband decomposition}
Interferences such as mains power and violin mode harmonics are
distributed over a large dynamic scale (the first harmonics are of
much larger amplitude than those of high order). But, since the
interferometer noise curve also decreases at low frequencies, their
relative amplitude as compared to the background noise power spectrum
at the same frequency remains large. Therefore, the model introduced
previously, namely that of large amplitude sinusoidal signals embedded
in broadband noise, is a reasonable approximation within the relevant
small bandwidth of frequencies.

For this reason, we divide the frequency axis in $p$ disjoint
frequency subbands of the same size. The $p$ signals lying in each
of the subbands are heterodyned and decimated to the sampling frequency
$f_s^{\textrm{band}}:=f_s/p$.

The tiling has the advantage that, if $p$ is sufficiently large, we
can consider the interferometer background noise almost white within a
subband, which implies that the noise has vanishing correlation time.
The prediction depth $d$ which has to be larger than the correlation
time, can be then simply fixed to any value greater than $1$ sample
period in each of the subbands.

\subsection{Long-term sinusoid removal}
Certain parts of the spectrum may not contain any long-term periodic
interferences. We apply a preliminary test to exclude subbands which
may not require the first denoising step. The test is crudely done by
estimating the amplitude $A$ of the sinusoid from the largest peak of
the power spectrum (Welch estimate) and comparing it to the variance
$\sigma^2$ of the broadband noise (also estimated from the power
spectrum). If it is found that $A>\sigma$, we decide that there exists
a long-term sinusoidal signal of sufficient amplitude in the band
which needs to be removed, otherwise we proceed directly to the second
step.

We apply the ALE in each of the selected subbands choosing parameters 
as follows:
\begin{itemize}
\item \textbf{Number of tap-weight coefficients $N$}\\  
  The number of tap-weight coefficients is fixed by prescribing an
  upperbound $0<\eta_{noise}<1$ to the ratio between the noise power
  corrupting the filtered output $y_k$ of the optimal filter
$\vec{w^*}$ and the input noise power. 
Let $\vec{n}_k:=(n_{k-d-m}, m =  0, 1,\ldots, N-1)^t$ 
be a collection of noise samples, then the above  condition reads~,
  \begin{equation}
    \E[(\vec{w}^{*t}\vec{n}_k)^2]\leq\eta_{noise}\E[(n_{k})^2],
  \end{equation}
  which, with the stationarity and whiteness of the background noise
  $n_k$, results in bound on the optimal filter gain~:
  \begin{equation}
    \vec{w}^{*t}\vec{w}^{*}\leq\eta_{noise}.
  \end{equation}

  The $L^2$ norm
  $\norm{\vec{w}^{*}}_2^2=(2/N^2)(N+\beta\cos(\gamma+2d\d))$ is
  obtained by squaring and summing the Eq. (\ref{wstar}) for the
  optimal filter. Since in typical cases $\beta \ll N$, this leads to
  simpler expression  $\norm{\vec{w}^{*}}^2_2\approx 2/N$.
  
  Consequently, the number of tap-weight coefficients $N$ has to be
  chosen so that,
  \begin{equation}
    \label{step1N}
    N \geq 2/\eta_{noise}.
  \end{equation}
\item \textbf{Step gain parameter $\mu$}\\
  We fix the step gain parameter by imposing to the distance of the
  ALE filter from optimality in the steady state to be smaller than a
  given threshold on average. As we have seen in Sect.
  \ref{steadystate}, this can be done naturally by imposing an
  upperbound $0<\eta_{sig}<1$ on the excess square mean error as
  compared to the signal power $E_s=A^2/2$~:
  \begin{equation}
    \xi^{(st)}\leq \eta_{sig} E_s.
  \end{equation}

  Using the expression obtained in the steady state analysis in Eq.
  (\ref{emse}) for the EMSE, this condition reduces to~:
  \begin{equation}
    \label{step1mu}
    \mu \leq \eta_{sig}/(N\sigma^2).
  \end{equation}
  
  Generally, this equation leads to small values of $\mu$ which prevent
  the convergence of the ALE filter from its initial state (i.e., all
  tap-weight coefficients are fixed to $0$) in a reasonable time
  (convergence faster than a tenth of second, which is the duration of
  the chunk of data). We solve this problem by first applying the ALE on
  a sequence of training data, the step gain parameter being set at the
  beginning to a large value (for fast convergence) and decreased
  gradually to the value given in Eq. (\ref{step1mu}). The filter
  obtained after the completion of this training is close to the 
  objective (i.e., the Wiener filter). We then start the 
  longterm sinusoid removal using this prepared filter.
  
  We remark here that although $\mu$ is small, it is \textit{non-zero} thus
  giving the ALE filter some flexibility of adapting to changes
  (non-stationarities) in the signal such as slow drifts in frequency
  and amplitude modulation. This property, however, needs to be 
  investigated more in detail.
\end{itemize}

\subsection{Ringdown removal}
\label{ringdownremoval}
The aim of the second step of the algorithm is to remove
oscillatory transients (ringdowns) of large amplitude. These
transients are either frequency bands excited from time to time
(caused by dysfunctions in the interferometer) or relics from the
previous step (when the envelope of a long-term sinusoid possesses fast
variations to which the algorithm cannot adapt or converge to during
the first step of removal).

The cleaning procedure consists in applying ALE the second time to 
each of the subbands but now, the parameters are so adjusted that, 
(\textit{i}) they select features with a larger bandwidth than in the
previous step, and (\textit{ii}) converge rapidly onto an
oscillatory noisy signal that may appear.

\begin{itemize}
\item \textbf{Number of tap-weight coefficients $N$}\\
  The impulse response duration and frequency selectivity (i.e., the
  filter bandwidth $\Delta f$) of the transfer function are dual in 
  character. This follows from the uncertainty relation. The rough 
  approximate relation between these quantities is given by,
  \begin{equation}
    \label{step2N}
    N=f_s/(p\Delta f),
  \end{equation}
  where $f_s$ is the sampling frequency. We choose the number of
  tap-weight coefficients $N$ by imposing a minimum
  bandwidth $\Delta f_{min}$ to the filter and using the above equation.
\item \textbf{Step gain parameter $\mu$}\\
  Assuming that the ringdown can be locally approximated by a sinusoid, we
  choose the step gain parameter by imposing a convergence time of the
  order of a typical transient duration (i.e., $t_{lock}\approx
  N\delta$). More concretely, setting $\rho:=\mu N A^2/2$ in the unnormalized
  form of Eq. (\ref{tlock}) (i.e., for arbitrary ringdown amplitude
  $A$), we solve for
  \begin{equation}
  \label{step2mu} 
  \frac{\ln(2\sigma^2 \rho)}{2\ln(1-\rho)}=N.
  \end{equation}
  
  Using the crude estimate $A^2/2 \approx \norm{\vec{x}_k}_2^2/N$ for
  the ringdown amplitude, the step gain parameter is finally obtained
  as $\mu=\rho/\norm{\vec{x}_k}_2^2$.
\end{itemize}

Since the ringdown signals are of short duration and can occur with
large time gaps, the ALE does not need to operate on each data
segment.  Accordingly, we have added a supervision test which decides
whether or not the denoising algorithm should be applied to a given
data segment. The test consists of observing the Gaussianity of the
filtered output $y_k=\vec{w}_k^t \vec{x}_k$. If the input signal $x_k$
is a zero-mean white Gaussian process of variance $\sigma^2$, then the
output of the filter $y_k$ shares the same characteristics, except
that the variance gets multiplied by the filter gain~: $\var y_k =
\norm{\vec{w}_k}^2_2 \sigma^2$. Furthermore, under this hypothesis,
the envelope ${\cal Y}_k=\abs{{\cal H}(y)_k}^2$ (${\cal H}$ denotes
the discrete Hilbert transform \footnote{The discrete Hilbert
  transform $y_n={\cal H}(x)_n$ of a signal $x_n$ is essentially
  obtained by cancelling its negative frequencies; more precisely,
  $Y(f):=2U(f)X(f)$, with $U(f)=1$ when $f\in[0,1/2]$ and $0$ when
  $f\in]-1/2,0[$ and where $X(f)$ (and $Y(f)$) denotes the Fourier
  transform of the corresponding signal
  $X(f)=\sum_{n=0}^{N}x_n\,e^{-2\pi i n f}$.} of $y_k$) follows by
definition a chi-square distribution with $1$ degree of freedom.

This implies that, up to an arbitrary probability $P_0$, 
the envelope ${\cal Y}_k$ does not exceed the threshold given below:
\begin{equation}
\label{envel}
{\cal Y}_k < \kappa(P_0) \,2 f_s^{band}\norm{\vec{w}_k}_2^2 \sigma^2,
\end{equation}
where $\kappa(\cdot)$ is the inverse function of the (unit variance)
$\chi^2$ cumulative distribution function (cdf).

If Eq. (\ref{envel}) is satisfied, we conclude that the filtered output
is essentially due to a Gaussian background noise and we leave the
input signal as it is. Otherwise, we conclude that the filtered output
carries a ringdown signal and decide to remove it from the input data.

The functioning of the second step of the denoising algorithm could be
interpreted as follows~: it removes from the input data, regions in
the time-frequency plane presumably associated with transients, whose
support is defined along the frequency axis by the ALE filter, and
along the time axis by the supervision criterion (\ref{envel}).

\vspace{5mm} 
After completing these two steps, we recombine the signal in all the
subbands together to retrieve a single strain signal.

\section{Numerical results}
\label{results}
\subsection{Simulated data: test of the ringdown removal} 
In this section, the goal is to test how effectively the second stage of 
the denoising algorithm (i.e., the ringdown removal) described in 
Sect. \ref{aleinpractise} operates on a simple signal.  The test
signal is composed of three ringdown signals (of fixed amplitude and
frequency) occurring successively in the data stream and embedded in a
additive Gaussian white noise. This model may be used to represent
ringdown disturbances originating from the same underlying physical
mechanism.

Each of these ringdown signals is a sinusoidal waveform, similar to
Eq.  (\ref{signaltest}) (with $A=1$, $f_0=50$ Hz and sampling
frequency $f_s=200$ Hz), whose support is limited in time by a
Gaussian envelope~:
\begin{equation}
r_k:=  A \exp\left(-\pi (t_k-t_c)^2/T^2\right)\,\c (2 \pi f_0 t_k+\Phi),
\end{equation}
where three different reference times $t_c$ are given and the
equivalent time duration is $T=200$ms (giving a frequency bandwidth
of $\Delta f\approx 1/T=5$ Hz and $Q:=f_0 T \approx 10$ cycles).

\begin{figure}
  \centerline{\includegraphics[width=110mm,height=90mm]{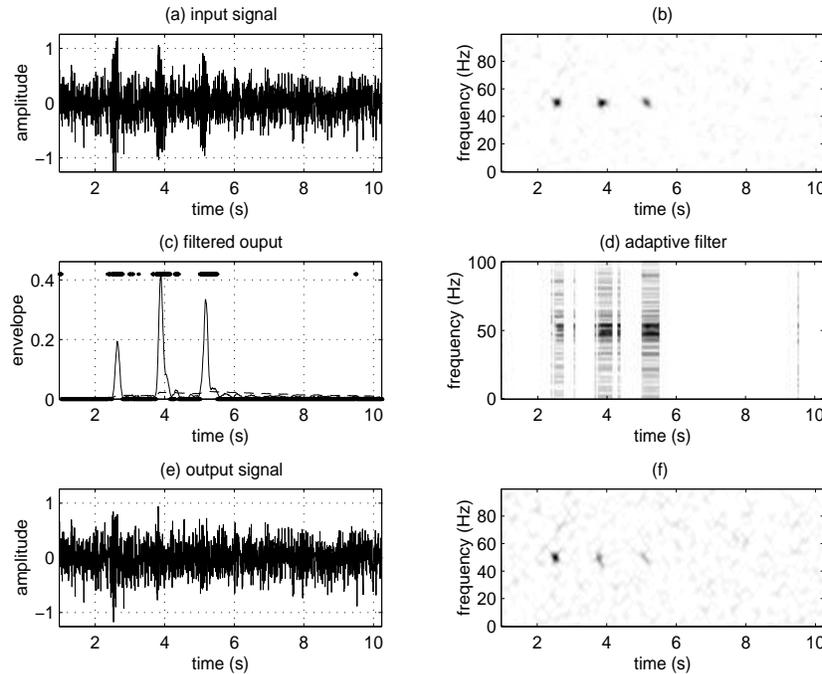}}
  \caption{\label{test2} \textbf{Applying the ALE to oscillating
      transients~: testing the ringdown removal algorithm}.  The 
    figure depicts the results of the transient
    removal algorithm presented in Sect.
    \protect\ref{ringdownremoval} to a signal in (\textbf{a}) composed
    of three successive oscillating bursts (see text for details)
    embedded in Gaussian white noise (SNR$=A^2/(2\sigma^2)= 8 (9
    \mathsf{dB})$).  In addition to the ALE, we measure the deviations
    of the filtered output from Gaussianity~: when its envelope 
    (\textbf{c}) exceeds some threshold (dashed line, see Eq.
    (\protect\ref{envel})), we decide that the filtered output is not
    normally distributed and, therefore contains a transient which has
    to be removed (this is indicated by dots at the top of the graph).
    The final net effect of the operation is that of time variant
    filtering of the input data. The corresponding transfer function
    is represented in (\textbf{d})~: the regions indicated with dark
    colours are parts of the time-frequency plane where the data are
    selected and removed from the input (i.e., it corresponds to the
    time-frequency ``band'' pass of the filter). Comparing the
    spectrograms \protect\cite{tftb} (see Eq. (\protect\ref{spectro}) for a
    definition) in (\textbf{b}) and (\textbf{f}) respectively of the
    input and output (\textbf{e}) signals, we observe that the three
    transients are progressively removed from the input (dark regions 
    represent large values of the time-frequency energy density).}  
\end{figure}

Figure \ref{test2} describes the application of the denoising
algorithm configured with $d=5$ sampling periods (equal to $25$ ms)
$\Delta f_{min}=3$ Hz, and $P_0=0.01$.  It can be seen that the
algorithm operates better on the transient encountered later in the
data train than its predecessor.  The explanation is that a transient
duration is too short for the filter to reach the steady state but,
when it encounters the next transient, the filter benefits from the
distance to $\vec{w}^*$ previously covered, thus improving the
convergence towards optimality.

This can be verified with a time-frequency representation
\cite{flandrin99:tf_ts} of the output signal such as Fig.~\ref{test2},
where we have chosen the spectrogram $S_x^h[n,m]:=|F_x^h[n,m]|^{2}$
defined as the squared modulus of the short-time Fourier transform~:
\begin{equation}
\label{spectro}
F_x^h[n,m]:=\sum_k{x_n h_{k-n}e^{-2\pi i nm}},
\end{equation}
where $n\in[1,2,\ldots, N]$, $m\in]-1/2 \ldots 1/2]$ and $h_k$ is an
arbitrary window (a Gaussian window here).

Notice that real time and frequency coordinates can be retrieved
through the relations~: $t=n/f_s$ and $f=m f_s$.

\subsection{Results on Caltech 40m proto-type data} 
Here we have applied the algorithm to the Caltech 40meter proto-type data
taken in October 1994 \cite{abramovici96:_improv_ligo}. This data was 
recorded with a sampling frequency of $f_s=9.86$kHz. We have used the
calibrated strain signal \cite{grasp} (relative arm length
measurement) for applying our algorithm.

We tile the complete spectrum into $p=32$ frequency subbands of
approximately $154$ Hz each. Each subband encounters typically one or
two long-term sinusoidal interferences.

We have chosen the prediction depth to be $d=5$ sampling periods, which
corresponds to a delay of $pd/f_s\approx 16$ ms in real time. The
correlation time of the broadband noise is effectively smaller in each
subband except at the extremities of the spectrum where the steep
slope of the spectrum does not allow us to assume the background noise 
to be locally white. It only affects the first and last subbands which are
not too important for detection purposes.

In the first stage, we have chosen $\eta_{noise}=0.01$ (giving $N=200$
according to Eq.  (\ref{step1N})) and $\eta_{sig}=0.01$. In the second
stage of ringdown removal, the minimum filter bandwidth has been fixed
to $\Delta f_{min}=3$ Hz, which gives a filter with $N=100$ tap-weight
coefficients (see Eq.  (\ref{step2N})) and we have set $P_0=0.01$ for
the Gaussianity test.

We have performed two types of simulations:
\begin{itemize}
\item a ``Caltech signal only'' simulation to measure improvements after
  denoising~: we check firstly, whether the frequency peaks are removed 
  from the noise power spectrum and secondly, whether the noise 
  statistics is closer to Gaussian than before denoising,
\item a ``Caltech+inspiral'' simulation to evaluate the
  consequences of the denoising algorithm on gravitational wave
  detection; specifically, for the case of the inspiralling compact
  binary signal. The question here is to check whether the denoising
  operation has removed a significant part or even whole of the
  inspiral signal.
\end{itemize}

\vspace{5mm} 
\noindent \textbf{Caltech signal only ---} Eleven of the
thirty-two frequency subbands (\# $1$--$9$, $11$ and $17$) are
selected and sent to the first cleaning step of the algorithm. In
these subbands, we obtained the following mean values for $A \approx
1.5 \times 10^{-16}$ and $\sigma \approx 3.6 \times 10^{-17}$ (the
sinusoid amplitude $A$ equals approximately $1$ to at most $5$ times
the noise standard deviation $\sigma$) leading to typical values for
the signal-to-noise ratio of about
$\mbox{SNR}=A^2/(2\sigma^2)\approx 8.7 (9.4 \mathsf{dB})$ and for the
step gain parameter (see Eq.  (\ref{step1mu})) of $\mu N A^2/2 \approx
0.04$ (spanning from $0.01$ to $0.14$).

The complete set of subband signals is processed in the second step.
The typical noise variance estimate is $\sigma=1.35 \times 10^{-17}$
(from $4.8 \times 10^{-18}$ to $10^{-16}$) leading according to Eq.
(\ref{step2mu}) to values of $\mu N \sigma^2$ which span the range 
of values from $0.07$ to $10^{-4}$.

\begin{figure}
  \centerline{ \begin{tabular}{c}
      \includegraphics[width=110mm]{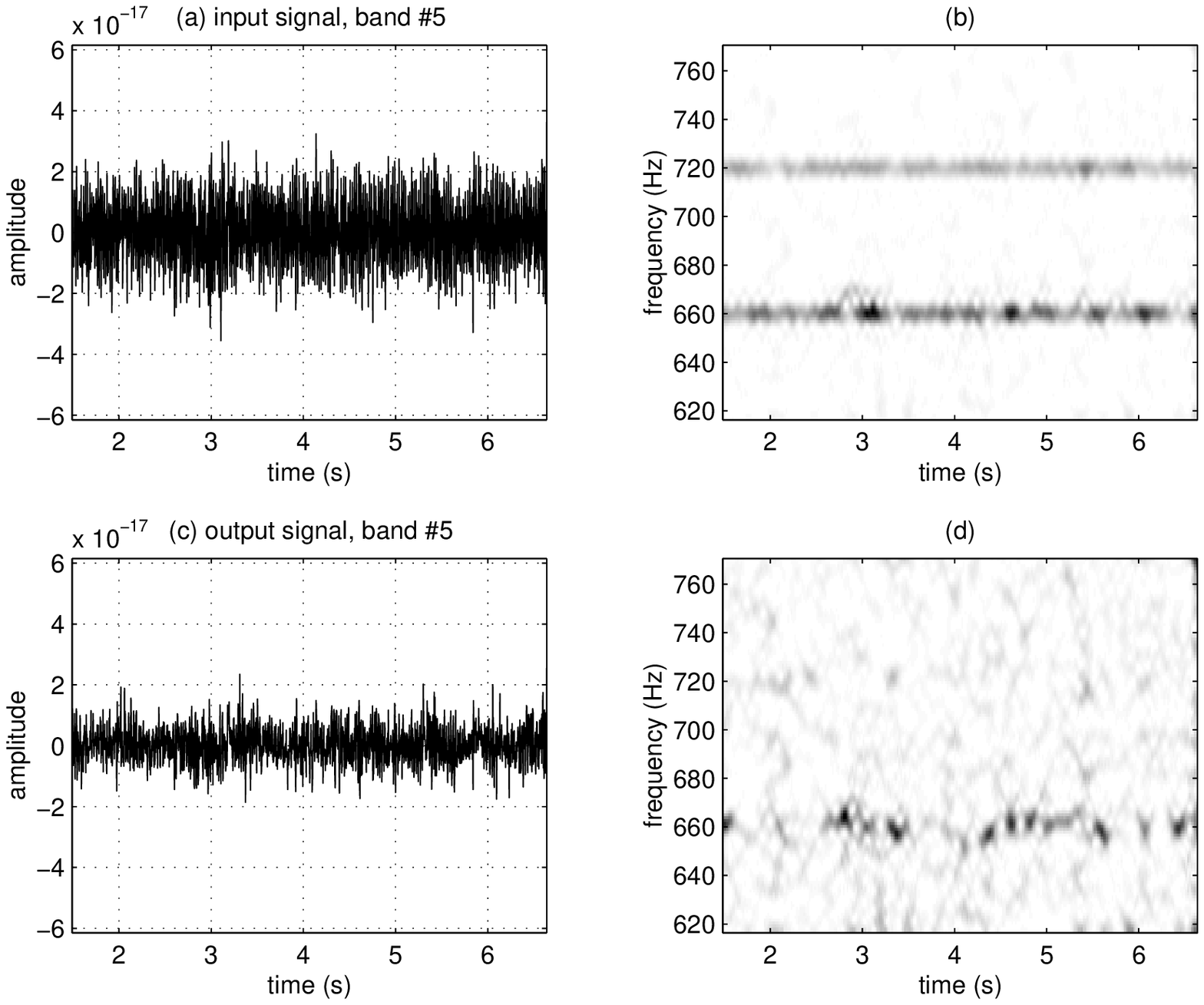}\\
      \includegraphics[width=110mm]{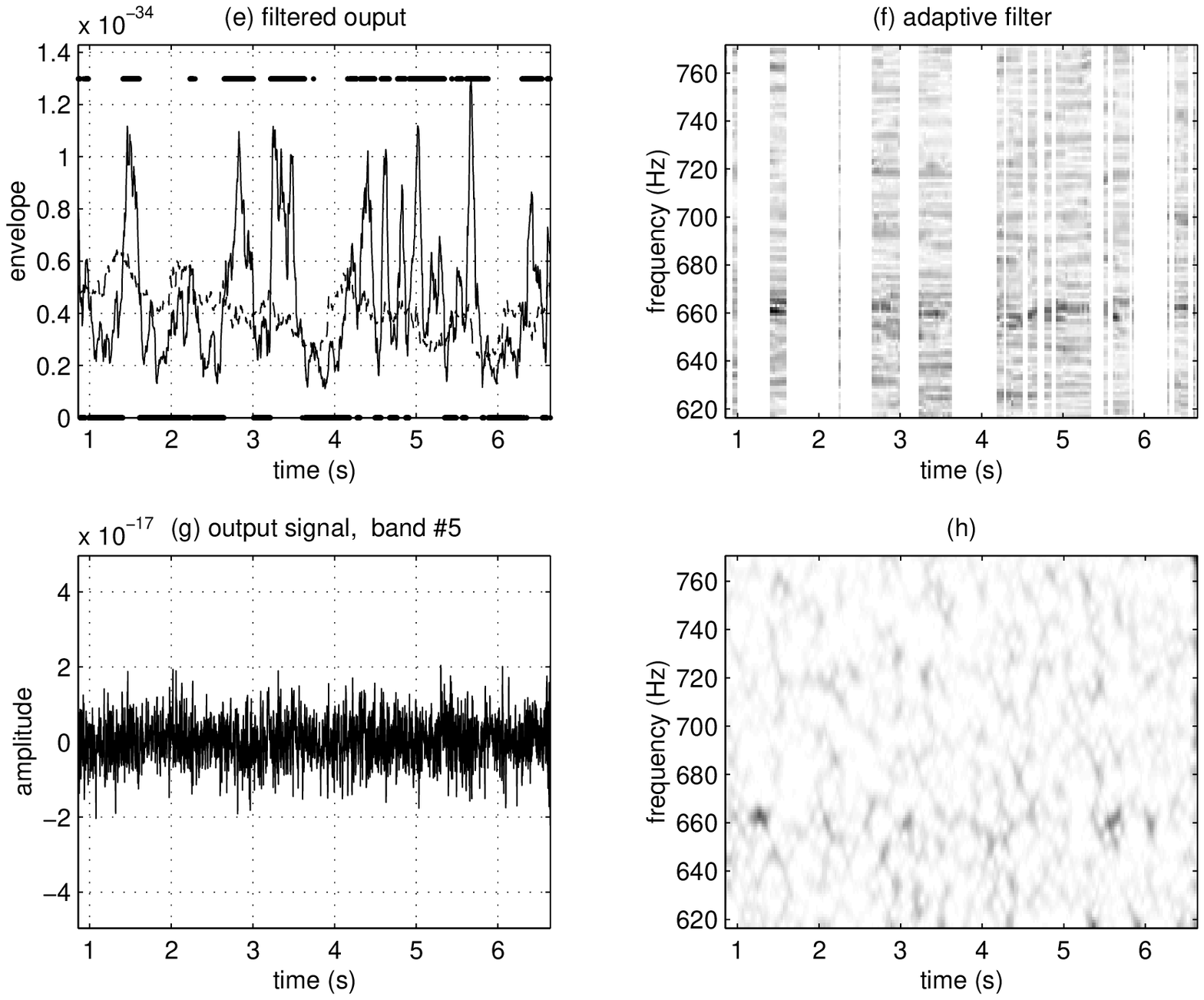} \end{tabular}}
  \caption{\label{stepfig} \textbf{Illustration of the denoising
      procedure on Caltech proto-type data}.  In the subband \#5
    (between 617 Hz and 771 Hz), the signal~\protect\cite{grasp} (the
    data were taken on the October, 14th 1994, frame \#2) in
    (\textbf{a}) contains two power line harmonics (at 660 Hz and 720
    Hz), which we are seen as darkened horizontal lines in the 
    spectrogram (see Eq. (\protect\ref{spectro})) in (\textbf{b}).  
    We apply ALE the first time to suppress long-term components 
    (\textbf{c}) with corresponding spectrogram (\textbf{d}).  
    A second run (\textbf{g})
    supervised by the criterion (\protect\ref{envel}) detailed in
    (\textbf{e}) and (\textbf{f}) (see the similar plots in Fig.
    \protect\ref{test2}, (\textbf{c}) and (\textbf{d}) for explanation)
    eliminates artefacts of shorter duration (such as fast
    fluctuations in the harmonic envelope).  Note that the spectrogram
    (\textbf{h}) of the final signal presents a homogeneous energy
    density both in time and frequency directions.}
\end{figure}

Figure \ref{stepfig} illustrates how the algorithm operates in the
fifth frequency subband (from $617$ Hz to $771$ Hz) among the $p=32$
ones being processed. This frequency band contains two power line
harmonics (the $11$th at $660$ Hz and the $12$th at $720$ Hz).

\begin{figure}
  \centerline{\includegraphics[width=110mm]{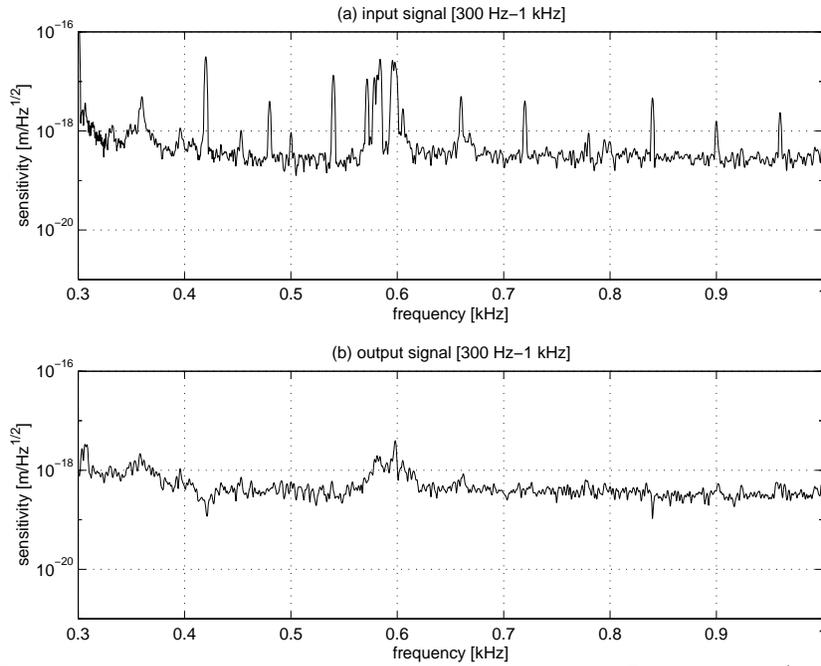}}
  \caption{\label{pspec} \textbf{``Caltech signal only''~: comparison
      between power spectra of ALE input/output signals}.  The figure
    depicts power spectra of the Caltech 40 meter signal (top) in
    the operating frequency band, between 300 Hz and 1kHz and the same
    signal after denoising (bottom).}  
\end{figure}

\begin{figure}
  \centerline{\includegraphics[width=110mm]{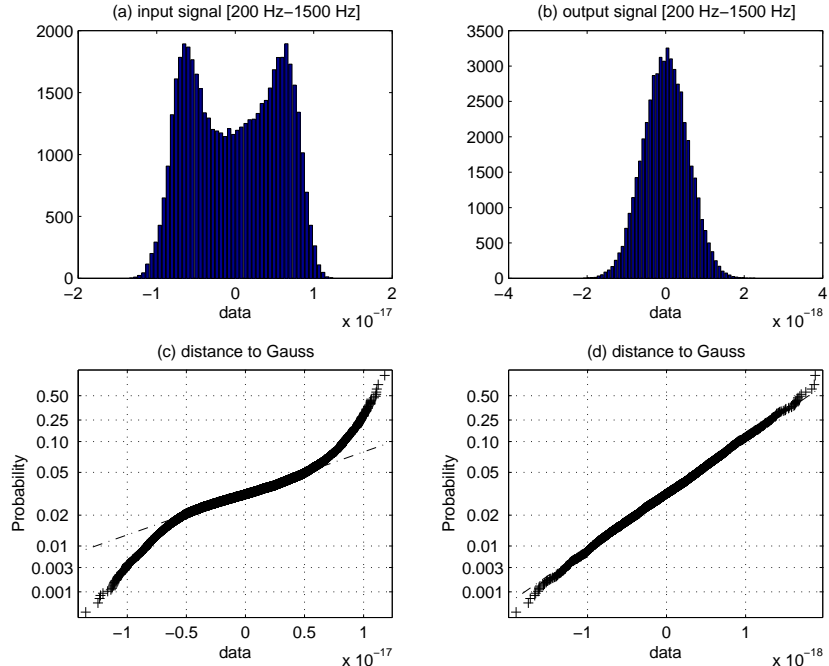}} 
  \caption{\label{hist} \textbf{``Caltech signal only''~: comparison 
      between histograms of ALE input/output signals}. The probability
    density functions of the Caltech 40 meter signal selected between
    300 Hz and 1kHz (left column) and the same signal after denoising
    (right column) have been estimated with histograms (top row). The
    bottom row shows the same histograms in special
    axes where a Gaussian bell curve appears as a straight line.}  
\end{figure}

Figures \ref{pspec} and \ref{hist} show respectively comparisons
between the power spectra and histograms of the signal before and
after denoising. We observe that after denoising, the frequency peaks 
have been removed from the input signal and the histogram appears much 
closer to the Gaussian bell curve.

\vspace{5mm} 
\noindent \textbf{Caltech signal + inspiral waveform ---}
The purpose of this test is to evaluate how the cleaning operation
affects gravitational wave detection and in particular to make sure
whether a significant part of the gravitational signature could be
removed from data. Answering this question by analytical means is
difficult, however a qualitative rational in the case of inspiral
binaries can be made and verified with simulations.

The theory predicts \cite{thorne87:_gravit} that the gravitational
waves emitted from inspiralling binaries of neutron stars are oscillating
waveforms whose frequency evolves in time in a prescribed manner and 
scans the interferometer bandwidth from lower end to the higher.

Their weak amplitude and short time duration within a single subband
(in the case we have considered, less than a second) make them
``invisible'' to the ALE filter. The amplitude and the duration of the
gravitational wave signal are simply not large enough for the ALE
coefficients to converge onto the gravitational wave instantaneous
frequency.

We have checked the validity of this argument by adding to the Caltech
signal the inspiralling `chirp' waveform in the Newtonian approximation 
\cite{thorne87:_gravit} of two  neutron star binaries each having a 
mass of $M=1.4$ solar masses, and located at a distance of 
$r=7\,$kpc from the Earth.

\begin{figure}
  \centerline{\includegraphics[width=110mm]{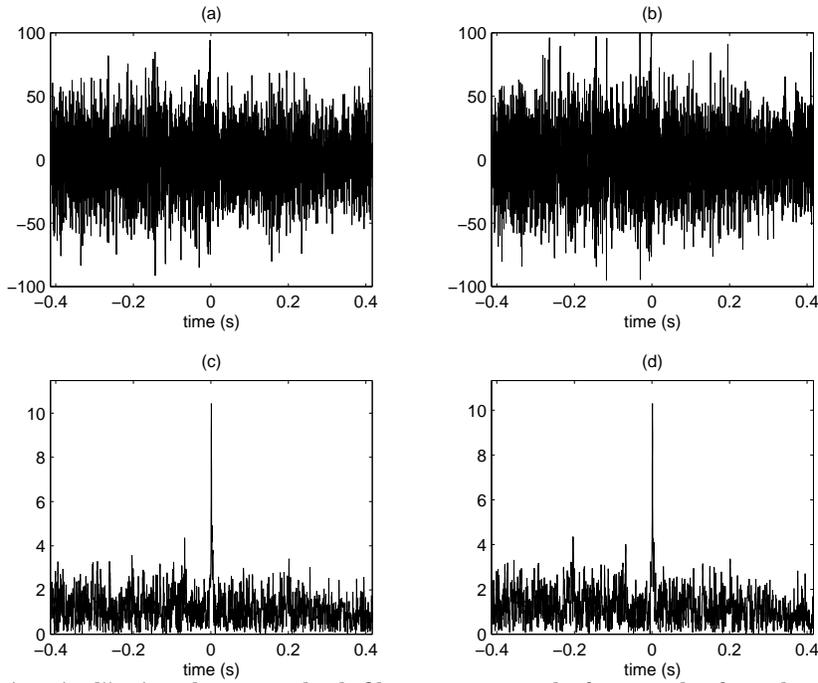}}
  \caption{\label{mafi} \textbf{``Caltech+inspiral'' signal~: 
      matched filter response before and after denoising}.  The
    Newtonian approximation of a gravitational wave emitted from an
    inspiraling binary (each with mass $1.4$ solar masses,at a distance of
    $7$ kpc and coalescence time fixed to $t = 0$) has been added to
    the Caltech interferometric proto-type data.  Top row plots show
    this signal (\textbf{a}) and its corresponding version after
    denoising (\textbf{b}) which have been selected and whitened
    within the frequency band $200$ Hz (i.e., the lower 
    frequency bound of the observation window) to $1.3$ kHz (i.e.,
    the predicted frequency for the last stable circular orbit of the
    binary).  The matched filter technique applied to detect the
    inspiral waveform, shows in both cases (e.g., without (\textbf{c})
    and with (\textbf{d}) denoising) a peak at time $t = 0$ in their
    detector responses (the normalization so chosen that 
    ``noise only'' detector fluctuations are of unit variance).}
\end{figure}

\begin{figure}
  \centerline{\includegraphics[width=70mm]{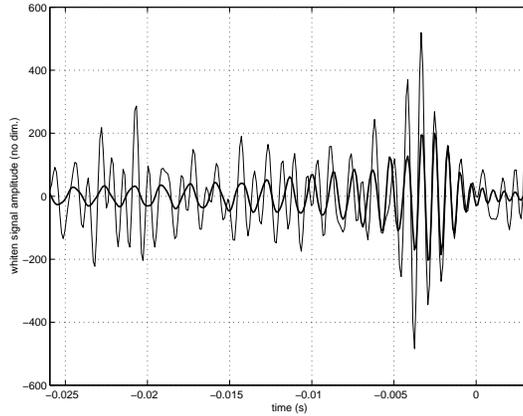}}
  \caption{\label{waveform} \textbf{``Caltech+inspiral'' signal~:
      zoomed view after denoising}. As an additional check, this
    diagram presents a zoomed view at the coalescence time $t = 0$ of
    the signal in Fig. \protect{\ref{mafi}}-\textbf{b} (whitened
    signal after denoising). We have superimposed on it (in bold) the
    signal as it would appear in the noise free case.}
\end{figure}

Figure \ref{mafi} depicts a comparison of matched filter detector
response on the same signal with and without denoising. The detector
output displays a peak of the same height and at the correct instant,
showing that the cleaning algorithm has not removed the inspiral
signal from the data. This can be crosschecked in Fig. \ref{waveform}
showing a zoomed view of the same signal after denoising.

\section{Concluding remarks}
The originality of the idea of the proposed denoising algorithm lies in its 
wide applicability, so that both types of disturbances, long-term 
sinusoidal and oscillatory transients (the type of noise which has
been ignored till now) can be treated. Although
the question of the computational burden in applying this algorithm
has not quite been addressed here, it appears from the simplicity of 
the operations involved (e.g., no requirement such as long-term FFTs)
that the total computational cost should be within acceptable limits, 
so that the algorithm can be operated in real time.
Furthermore, the structure of the algorithm already implemented with Matlab
\cite{ale} can be easily translated into a parallel code (each
processing node can be associated with one frequency subband and the 
processing can be done independently).

As part of future extensions to the present work, some improvements to
the current code might be needed~: in order to limit the finite size
effects in the subband decomposition and reconstruction, a reversible
filter bank (e.g., a Gabor transform) would be preferable than the
crude method used here.

The key idea (i.e., looking for correlation between the current sample
of the strain signal and a reference signal, namely a set of past
samples) can be also extended to investigate correlations of the
detector output with other environmental channels by simply using them
as a reference rather than the strain signal itself. Similarly to the
cross-talk removal in \cite{allen99:_autom} but with adaptive methods,
such an algorithm would provide an estimation of any poorly known
(linear) transfer functions relating noise sources to their final
leaking in the detector output and of the environmental contamination
that must be subtracted from the data, if so desired.

\subsection*{Acknowledgments}
We would like to thank B. F. Schutz for suggesting the idea of
adaptive methods and also for fruitful conversations and the LIGO
collaboration for providing us the Caltech 40meter proto-type data. E.
C.-M. would like to thank W.  Anderson, R. Balasubramian, J.
Creighton and S. Mohanty for their useful comments and suggestions.

\clearpage
\newpage
\onecolumn

\end{document}